\documentclass[conference]{IEEEtran}
\IEEEoverridecommandlockouts
\usepackage{mathtools}
\usepackage{cite}
\usepackage{amsmath,amssymb,amsfonts}
\usepackage{algorithmic}
\usepackage{graphicx}
\usepackage{textcomp}
\usepackage{xcolor}
\usepackage{lipsum}
\usepackage{cuted}
\usepackage{array}
\usepackage{comment}
\usepackage{url}
\def\BibTeX{{\rm B\kern-.05em{\sc i\kern-.025em b}\kern-.08em
    T\kern-.1667em\lower.7ex\hbox{E}\kern-.125emX}}
\begin{document}

\title{Co-optimization of Operational Unit Commitment and Reserve Power Scheduling for Modern Grid}

\author{Ramkrishna Mishan, Matthew Egan, Mohammed Ben-Idris, and Hanif Livani  \\ Department of Electrical \& Biomedical Engineering, University of Nevada, Reno, Reno, NV 89557 
\\(emails: \{mishan, matthewegan\}@nevada.unr.edu, mbenidris@unr.edu, and hlivani@unr.edu )\vspace{-0.2ex}}

\maketitle

\begin{abstract}

Modern power grids combine conventional generators with distributed energy resource (DER) generators in response to concerns over climate change and long-term energy security. Due to the intermittent nature of DERs, different types of energy storage devices (ESDs) must be installed to minimize unit commitment problems and accommodate spinning reserve power. ESDs have operational and resource constraints, such as charge and discharge rates or maximum and minimum state of charge (SoC). This paper proposes a linear programming (LP) optimization framework to maximize the unit-committed power for a specific optimum spinning reserve power for a particular power grid. Using this optimization framework, we also determine the total dispatchable power, non-dispatchable power, spinning reserve power, and arbitrage power using DER and ESD resource constraints. To describe the ESD and DER constraints, this paper evaluates several factors: availability, dispatchability, non-dispatchability, spinning reserve, and arbitrage factor. These factors are used as constraints in this LP optimization to determine the total optimal reserve power from the existing DERs. The proposed optimization framework maximizes the ratio of dispatchable to non-dispatchable power to minimize unit commitment problems within a specific range of spinning reserve power set to each DER. This optimization framework is implemented in the modified IEEE 34-bus distribution system, adding ten DERs in ten different buses to verify its efficacy.
\end{abstract}


\section{Introduction}

Different types of renewable distributed energy resources (DERs) including solar photovoltaic panels/plants and wind turbines integrate with traditional generators (Steam Turbines, Gas Turbines, Diesel Generators, etc.) in the modern grid's transmission and distribution power network. 
Under high renewable energy penetration, DERs should be utilized to provide unit commitment and spinning reserve power alongside conventional generators \cite{initiative2012managing}. However, the intermittent nature of DER generators creates a unit commitment problem\cite{7592421}. Also, if the DERs are assigned to provide reserve power without Energy Storage Devices (ESDs), the DERs may operate at set points below the available renewable power. In such a case, aside from non-committed reserve power, some portion of renewable energy is wasted. To make DERs responsible for stable grid operation,  the ESDs incorporate these distributed generators \cite{denholm2010role}. Without using a proportionate amount of ESD size, the high penetration of DER into the existing grid creates grid power fluctuations, which need to be mitigated by optimizing the reserve power scheduling and minimizing the unit commitment problem of an individual or a cluster of DERs \cite{yang2018battery}. However, to generate the unit committed and reserve power from the combined solar-storage system, the renewable plant designer needs to explore all individual components (PV Array/Plants, Inverters, ESDs) resource and operational constraints or the combination of these components when operating concurrently. 

In\cite{arabali2016new}, the authors concentrated on grid-scale storage systems for day-ahead optimal scheduling. However, the authors assumed that storage devices' reserve power uniformity could not be maintained due to limitations of discharge rates or ramp rates. Although in \cite{wagner1997large} the authors suggested a large storage system for load labeling when integrating with solar power application, they overlooked how the charging/discharging cycle of storage systems significantly reconfigure the active power reserve market. By using a proper expression of reserve power associated with various risk levels, a model for discovering optimal unit commitment power was developed in \cite{4181668}. In \cite{6038963}, the authors presented a single mixed-integer optimization model after efficiently integrating unit commitment and generation growth planning. Nonetheless, these papers did not explore all the resource and operational constraints when incorporating ESDs with PV inverters. Identifying the operational factor and designing operational constraints after considering all the DER resources is vital to extract unit-committed and spinning reserve power.   

Estimating unit committed renewable power generations with specific amount of reserve power is critical for a stable power grid. Unit commitment (UC) can be expressed as the "determination of generating units to be invariant during a short-term scheduling period (hours, a day, or a week)" \cite{abujarad2017recent}. The UC must satisfy system demand and reserve requirements in an optimal, cost-efficient manner.  Due to high variability in the demand side, the researchers propose several factors, such as the utilization factor and diversity factor \cite{remp1949fundamentals}. These factors are utilized to accurately schedule the generation power and reserve power for the conventional power plant. Distributed renewable power generation has a high variability like load demand power. Also, in the conventional generation system, the unit commitment decision is largely influenced by unit constraints, synchronization, start-up, shutdown, system capacity, spinning reserve, and non-spinning reserve \cite{morales2017hidden}. However, the DER inverters' start-up and shutdown capacity are not the problems; instead, their issue is intermittency. Due to the high variability of the DER generator, this paper applies several operational factors as a constraint to find the maximum unit commitment power. When a DER generator integrates with the storage device connected to the grid, a portion of its power is immediately scheduled based on available power and storage resource constraints. Also, a small segment is scheduled for reserve power, depending on the system operator's request or load demand profile. In this way, total scheduled power is flexible. Due to this flexible scheduling, DERs operate similarly to a dispatchable operational mode of conventional generators during the storage discharging mode.

 
Integrating the intermittent DER into the power grid would create frequency and voltage fluctuation due to the unit commitment problem. Therefore, before providing ancillary services like frequency or voltage regulation or spinning reserve, the DER must provide the unit committed power. 
Without any storage incorporation, the DER would always generate non-unit-committed or non-dispatchable power even with the perfect prediction of its generation. Therefore, it is critical for ISO operator to know the maximum possible extraction of the unit-committed power for a given DER resource and its operational constraints.
 
Before determining operational constraints, this paper defines several operational factors such as availability factor, dispatchability factor, non-dispatchability factor, spinning reserve factor, and storage arbitrage factor. One of the main contributions this paper provides is to differentiate different power profiles such as dispatchable/unit committed power, non-dispatchable power, arbitrage power, and reserve power when performing operational optimization. After that, this paper maximizes the unit committed power (UCP) considering all these resource and operational constraints for an optimal spinning reserve power. In the proposed optimization framework, the UCP ratio of DERs will increase for a set of optimal reserve power units by properly regulating ESDs discharging/charging cycle and accurate reading of the current state of charge (SoC). In different seasons and on different days, the output of solar PV can be higher or lower and necessitate higher ratios. In such cases, this ratio can be further enhanced after increasing the energy arbitrage factor for specific storage system. 

The paper is organized as follows: Section II covers several resource constraints and different operational factors. Section III discusses the model formulation of the optimization constraints and the objective function. Section IV presents the results of case studies.
 
\section{Resource Constraints and Different Operational Factors}
DER inverter without incorporating an ESD does not contribute to the grid unit commitment and reserve power \cite{howlader2020energy}. Even if we can predict the DER generation correctly, the unit commitment is not maintained because of its intermittent nature. If a DER is not connected to any storage system and the system operator sets some reserve power, that power might be wasted, which is not ideal. The DER power profiles remain in the black-box model until this point. When the DER owners join the electricity market, and the ISO operator selects a particular DER set based on prices, they are unaware of all the power profiles which reflect the power quality. In current models,  both DER owners and ISO operators rely heavily on the electricity price, which may not always consider in the best interest of a stable grid.

Along with dynamic demand power, the high penetration of DER into the electric grid causes the grid to be volatile because of the uncertainty in the dispatchable and spinning reserve amount. Determining the UCP and spinning reserve power (SRP) from the DERs enhance the grid stability or regulation resources. To assess the DER's resources and operational constraints, this paper proposes several operation factors such as DER's Availability Factor (AF), Dispatchability Factor (DF), Non-Dispatchability Factor (NDF), and Storage Arbitrage Factor (SAF). 

Essentially, these factors consider the combination of DER's components resource constraints. They accommodate the unit commitment both in DER set point and reserve power considering the operational and resource limitations. Calculating these factors simplify determining the total UCP and SRP for an individual DER, cluster of DERs, and total DERs. In addition, these factors accommodate to maximize the number of UCP within a specific interval. Initially, the arbitrage factor and discharge rate of the ESD are set by the system operator to know the maximum amount of dispatchable, non-dispatchable, spinning reserve and arbitrage power. Also, knowing these power profiles and maximum UCP will help the DER owner to join the electricity market suitably. This paper proposes the availability factor which calculates the total amount of solar PV and storage power in terms of the inverter's rated capacity.

Furthermore, the system operator determines the initial set-point of the spinning reserve factor considering the ESD' mode and uniformity of the reserve power. Later, the optimization algorithm determines the final setpoint of dispatchable, non-dispatchable, and reserve factor points. Finally, the dispatchability and non-dispatchability factors determine the number of unit-committed and non-dispatchable power. The size of DER integration increases over time; knowing the ratio of UCP or dispatchable power vs. non-UCP provides information regarding the grid power availability.  This paper also provides the standard size for inverter and storage devices to maximize the resources.

\subsection{DER Availability Factor (AF):}
This paper suggests the energy availability factor or simply availability factor in determining the total available power or energy of both PVs and ESDs. After assigning the specific interval discharge rate of the storage devices, the system operators need information regarding the total available power or energy. Considering the predicted solar PV and ESD generation compared to the inverter rated-power provide the total available energy. Determining this factor is effective in several ways. First, this factor informs the system operator regarding the portion of total available power ready to be dispatched and reserved during the ESD discharge mode. Also, this factor reports the total amount of non-dispatchable and arbitrage power during the storage charging mode. Using this factor improves ESD lifespan by not allowing ESD to operate beyond the maximum discharge rate and preventing to operate charging/discharging simultaneously. Finally, this factor maximizes the inverter life span by not allowing the inverter set-point to go beyond its rated capacity during the storage discharging mode.  As mentioned before, this factor is used in the optimization algorithm to determine percent of UCP and spinning reserve power from the total available power.

The ESD can connect to the grid in several points in the modern grid, such as small size in the Behind-the-meter (BTM) and relatively large utility-scale of the substation distribution feeder. Behind-the-meter ESDs represent the customer-sited stationary storage systems coupled to the distribution system and corporate customers' side utility service meter.

\subsubsection{DER Availability Factor for BTM' ESD}
Under this connection, the customer able to provide UCP during the storage discharging cyle.
For the BTM storage system, the availability factor would be 

\begin{equation}\label{equ:inv_AFF}
\begin{aligned}
    \mbox{AF}= \frac{\mbox{\%(SoC-SoC}_{m})\times ESC \times \frac{1}{DR}+P_{pv}^{pred} }{P_{inv}}
\end{aligned}
\end{equation} 
where AF stands for Availability Factor, $SoC_{m}$ is the minimum SoC, ESC is the Energy Storage Capacity of a given battery, DR is Discharge Rate in Hours, $P_{pv}^{pred}$ is the historical solar PV average output with in certain interval, measured in kW, and $P_{inv}$ is the DER inverter's rated capacity. $P_{pv}^{pred}$  is a rudimentary prediction of the amount of power generated at a given point, though it can also be measured in conjunction with things such as weather or condition of the solar panels.

This AF will determine the total energy availability by considering both battery energy and the average historical performance of solar PV. To correctly calculate the available energy of the battery, this factor includes the depth of discharge as ($SoC-SoC_{m}$) and discharging rate (DR) and inverter rated power. Further, this energy availability factor is a constraint when determining the Dispatchable Factor (DF) and Reserve Factor (RF).

\subsubsection{DER Availability Factor for Utility Scale' ESD} Under this condition, we assume only utility-scale storage devices connect to the distribution feeder or the substation. If no other BTM storage connect to aggregated n number of DER under this condition, then the AF would be,
\begin{equation}\label{equ:inv_AF_Ut}
\begin{aligned}
    \mbox{AF}= \frac{\mbox{\%(SoC-SoC}_{m})\times ESC\times \frac{1}{DR}+P_{pv1}^{avg}+\hdots+P_{pvn}^{avg} }{P_{inv1}+P_{inv2}+\hdots+P_{invn}+P_{USESD}}
\end{aligned}
\end{equation}
where $P_{USESD}$ is the utility scale inverter size. 


After determining the energy availability factor, the next step is determining the operational dispatchablility and spinning reserve  factor. This paper proposes the dispatchability factor or DF to determine total dispatchable power or UC power for a DER inverter.

\subsection{DER Dispatchability Factor:}

To determine the maximum available dispatchable or UCP from the total available power, we proposed the dispatchability factor as a percent of inverter rated power. This factor will only be used in battery storage discharging mode when incorporating a grid-forming DER inverter to minimize the power supply fluctuation. It refers to the fraction of rated inverter power that is dispatchable based on the power grid operator request. Dispatchable power is similar to constant power set-point, which will not be affected by the intermittent nature of solar radiation. Usually, this power operates above the DER historical average power as storage devices incorporate it.  Choosing a proper DF is crucial to minimize the PV generation losses. If dispatchable power is below the historical average of PV generation, some portion of the generation might waste as the battery is already in discharging mode, or it creates mini charging and discharging cycle if the SoC of storage devices is low.
\begin{equation}\label{equ:inv_DF}
\begin{aligned}
    DF= \frac{\mbox{ Dispatchable \ Power  }}{P_{inv}}
\end{aligned}
\end{equation}

\subsection{Spinning Reserve Factor:}

Spinning Reserve Factor (SRF) is only calculated whilst the energy storage system is discharging. This unit-less factor represents the individual DER inverter spinning reserve power, normalized for its rated power. Oftentimes, this reserve power is supplied by the storage system. However, if the system operator demands more reserve power than what is determined by this factor, that is more losses or mini charging/discharging cycles that affects battery life.

\begin{equation}\label{equ:inv_SRF}
\begin{aligned}
    SRF= \frac{\mbox{Spinning \ Reserve \ Power }}{P_{inv}}
\end{aligned}
\end{equation}

\subsection{Storage Arbitrage Factor:}
In the charging mode of the battery storage system, the DER inverter's setpoint is affected by the solar radiation and the energy arbitrage factor.
A certain amount of PV generation is consumed to charge the battery as it charges. The system operator usually sets this amount to store the charge during the off-peak hour and deliver in a peak-load hour. 
If the system operator does not set it, by default, the optimized ESD's charging rate determines this factor.

\begin{equation}\label{equ:inv_SAF1}
\begin{aligned}
      SAF= \frac{\mbox{Charging Rate}}{P_{inv}}
\end{aligned}
\end{equation}
where Charging Rate of the Storage Device in Kw.

\subsection{DER Non-dispatchability Factor:}

This factor will only be used if a PV inverter is connected without an ESD or ESD' SoC is too low to dispatch, or the ESD is in charging mode. In that case, the supplied power of the inverter is controlled by solar radiation and cannot give unit-committed power. This is implemented to account for the main weakness of renewable energy: that there is no way to control the power without shutting off the inverter.
\begin{equation}\label{equ:inv_SAF2}
\begin{aligned}
      NDF = \frac{\mbox{Non-dispatchable Power}}{P_{inv}}
\end{aligned}
\end{equation}

\subsection{PV Inverter Sizing for Behind-the-meter Storage Support:}
The inverter sizing can depend upon several factors such as average solar radiation, peak radiation, line capacity, and connected battery sizing, if any battery is installed. When the DER and ESS incorporate with the inverter, the system provides unit commitment and reserve power; the power operator wants to know total available power or energy considering the solar prediction, ESD' capacity, SoC,  and optimal discharge rate of a specific interval. This amount might cross the inverter rating if a low inverter rating is chosen. In this work, the inverter is sized to be equivalent to the rated PV power plus the rated storage discharge rate, plus a percentage in anticipation for of future upgrades.
\begin{equation}\label{equ:inv_SAF3}
\begin{aligned}
      P_{inv}= P_{pv,max}^{pred} + DR^{max}
\end{aligned}
\end{equation}
where $DR^{max}$ is the storage maximum discharge rate in kW.

\subsection{Grid-Connected BTM and Utility Scale Energy Storage:}
Energy storage system design can be categorized into two types: Centralized and Decentralized or Distributed \cite{zakeri2021centralized}. In centralized storage design, the ESDs provide the peak shaving \cite{zakeri2021centralized} but might need to provide high power density discharge. Also, the centralized storage system will face reliability issues as a few cell damage might cause a chain reaction and affect the whole system \cite{wu2019safety}. However, a decentralized or distributed storage prevents over-voltage on distribution feeders \cite{7754038} but supports frequency regulation \cite{6612760}.  Therefore, this paper considers that a decentralized storage system with DER generators will be the modern grid, primarily focusing on solar PV systems. 

The decentralized storage system creates a more robust system in terms of providing ancillary services  in distribution feeders, which provides frequency regulation, protect from overvoltage, and increase reliability \cite {6612760,7754038}. In addition, it increases reliability by providing power to communication devices in case of a DER shut down. 
In this paper, we specify the standard size of storage devices that increase DER UCP unit without losing power or creating a mini storage charging/discharging cycle. The proposed storage size (SS) in Ah is as follows,
\begin{equation}\label{equ:inv_SAF4}
\begin{aligned}
      SS = \frac{K_p\times PV \ Size \times DR^{max} }{SSV \times K_T \times \eta_{s} \times \eta_{CC} \times \eta_{w} \times DoD\times D_T  }
\end{aligned}
\end{equation}
where SSV is storage system DC voltage, $K_T$ is solar clearness index,  $\eta_{s}$ is storage system efficiency,  $\eta_{CC}$ is charge controller efficiency, $\eta_{w}$ is wiring efficiency, DoD is dept of discharge, $K_p$ is percent proportional depends upon the solar power quality and investors, $D_T$ is the temperature derating factor, and $DR^{max}$ is the storage maximum discharge rate is defined in hours.

\subsection{Storage Charging and Discharging Approaches:}

Charge controllers are used to manage the charging and discharging rate of ESDs to smooth PV generations from the inverters \cite{5616799}. Over-charging the battery may cause electrolysis, releasing H2 and O2 gases, thereby creating a fire hazard. This is avoided through each ESD's designated maximum charging limit. If the battery's discharge rate is too great, the battery will drained rapidly. Thus, the battery would cycle frequently enough to cause premature battery failure. A battery's internal resistance is proportional to its charge, therefore a battery nearer to full charge draws a lower charging current, and thus wastes less energy. At different charging levels, there are different rates of charging and discharging. So, multistage controllers can be used to provide a more efficient methods of charging and discharging, according to \cite{eldahab2016enhancing}. 

\section{Model Formulation: Optimization Objective Function and Constraints Design}\label{sec:con}\
This paper extracts the maximum available UCP from a specific distribution grid after considering the fixed minimum spinning reserve. Before extracting this total spinning reserve from these DERs, the algorithm sets the individual minimum and maximum spinning reserve for a particular DER. 

\subsection{Objective Function}

From the definition of DF, it is clear that larger DF would provide larger dispatchable or unit-committed power. The objective is formulated to maximize the UCP or dispatchable power for a given spinning reserve power of a distribution system as follows,
\begin{equation} 
\max \sum_{p=1}^{n} 
\left[ P_{inv,p}^{rated} \times DF_p \right] 
\end{equation}

The optimization of these allocated powers is crucial and maximizing storage devices' efficiency by properly regulating its charging, discharging cycle and accurate reading of the current SoC.
Furthermore, this optimization technique is critical to identify the total available energy (both in storage devices and the predicted power generation of solar PV)  compared to the inverter rated capacity. Subsequently, the allocated dispatchable and reserve power is determined. The summation of the dispatchable and reserve amounts shouldn't exceed the inverter rated power. 

DF and SRF are only used in the discharging cycle of DERs' ESD. If the dispatchable amount is set less, energy could be lost in the peak solar radiation as the battery is already set for discharging mode. In the case of lower SoC of battery, there could be a mini charging/discharging cycle. For this reason, this optimization technique maximizes the UCP for a given total reserve power and individual minimum SRF. 

By implementing all the constraints discussed below, the system operator controls the ESDs charging and discharging rate. 

\subsection{Energy Storage Constraints:}The SoC constraint for the battery is as follows,   
\begin{equation}\label{equ:batt_SoC1}
\begin{aligned}
     SoC_{min} 	\leq SoC 	\leq SoC_{max}
\end{aligned}
\end{equation}
where $SoC_{max}$  should be 100 for fresh batteries, but will typically decrease over time for units with memory problems such as lead acid or lithium ion batteries.


\begin{equation}\label{equ:batt_SoC2}
\begin{aligned}
      0 < DR	\leq DR^{max}	
\end{aligned}
\end{equation}
where $DR^{max}$ is the maximum discharge rate allowed by ESDs. Also the charging rate constraints are as follows,
\begin{equation}\label{equ:batt_SoC_C}
\begin{aligned}
    0 < CR 	\leq CR^{max}
\end{aligned}
\end{equation}
where $CR$ is the storage charging rate and $CR^{max}$ is the maximum charge rate allowed by ESDs. If DER owner charge the BTM' ESD from PV directly 
and $P_{pv}^{pred}\ (max)  \leq CR^{max}$, then charging rate constraints are as

\begin{equation}\label{equ:batt_SoC_CV}
\begin{aligned}
      0 < CR 	\leq P_{pv}^{pred} \ (max) 	
\end{aligned}
\end{equation}
which denotes that the maximum power taken from the system at a given instant is identical to the power the PV system generates, which cannot be controlled, plus the amount of power stored in the battery, which can be controlled. These constraints are intuitive and in line with real world systems.

\subsection{Dispatchable Factor Constraint: }

As stated in Section \ref{sec:con}, the DER only activates dispatch mode when the storage cycle is in the discharging mode. The minimum unit committed amount is ideally equal to the predicted solar power average, and the maximum value is determined in the optimization after assigning individual minimum spinning reserve factor and total reserve amount.   
\begin{equation}\label{equ: inverter_POS}
    \begin{cases}
    P_{pv,1}^{pred} \leq DF_1 \times P_{1}^{ inv} \leq AF_1 \times P_{1}^{ inv} \\
    P_{pv,2}^{pred} \leq DF_2 \times P_{2}^{ inv} \leq AF_2 \times P_{2}^{ inv}\\
     \vdots\\
    P_{pv,n}^{pred} \leq DF_n \times P_{n}^{ inv} \leq AF_n \times P_{n}^{ inv} 
    \end{cases}       
\end{equation}

\subsection{Spinning Reserve Power Constraints}
Spinning reserve operate only the discharging mode of ESDs. The system operator assigns the individual DER minimum reserve and total reserve power within certain limit as follows,
\begin{equation}\label{equ: inv_SRF}
    \begin{cases}
     SRF_1^{min} \leq SRF_1 \leq SRF_1^{max}  \\
     SRF_2^{min} \leq SRF_2 \leq SRF_2^{max}  \\
\vdots\\
    SRF_n^{min} \leq SRF_n \leq SRF_n^{max}
    \end{cases}       
\end{equation}

After determining the individual DER maximum and minimum setting, the system operator may define the total minimum spinning reserve for a particular distribution system by following operational constraints.

\begin{equation} 
\sum_{p=1}^{n} SRF_p^{min} \times P_{p}^{inv} \leq TSRP \leq \sum_{p=1}^{n} SRF_p^{max} \times P_{p}^{inv}
\end{equation}
where TSRP is the total spinning reserve power requirement from this distribution system. However, an optimal spinning reserve from a particular distribution system for each interval is fixed. Therefore, this power is constant for a single optimization loop considering the above spinning reserve power constraints.

\begin{table*}[h!]
    \centering
        \caption{Initial conditions of the system and Solar + Storage characteristics.}
\begin{tabular}{|c|c|c|c|c|c|c|c|c|c|c|c|}
\hline
\textbf{DER} & \textbf{$Bus$ (phases)} & \textbf{$P_{inv}$ (kW)} & \textbf{$P_{PV,his}$ (kW)} & \textbf{$Storage \ Cap.$ (kWh)} & \textbf{$SoC$ (\%)} & \textbf{$SoC_{min}$ (\%)} & \textbf{$AF$} & \textbf{$DF$} & \textbf{$SRF$} & \textbf{$SAF$} \\ \hline
DER1  & 890 (ABC)   & 60        & 45                             & 16                     & 65  & 30       & 0.96                      & 0.8                     & 0.10                        & 0.4              \\ \hline
DER2  & 844 (ABC)   & 75        & 60                             & 20                     & 85  & 20            & 1.03                      & 0.78                    & 0                          & 0.29             \\ \hline
DER3  & 860 (ABC)   & 90        & 81                             & 24                     & 50  & 20       & 0.98               & 0.88                    & 0.11                      & 0.49             \\ \hline
DER4  & 848 (ABC)   & 82.5      & 75                             & 22                     & 90  & 20         & 1.18                      & 0.9                     & 0.10                        & 0.52             \\ \hline
DER5  & 830 (C)    & 45        & 37.5                           & 12                     & 95  & 30            & 1.05                      & 0.79                    & 0                          & 0.52             \\ \hline
DER6  & 822 (A)     & 97.5      & 82.5                           & 26                     & 40  & 35         & 0.85               & 0.75                    & 0.10                        & 0.31             \\ \hline
DER7  & 806 (B)     & 75        & 66                             & 20                     & 80  & 35            & 1.03                     & 0.78                    & 0                          & 0.36             \\ \hline
DER8  & 836 (C)    & 52.5      & 42                             & 14                     & 72  & 35         & 0.92               & 0.85                    & 0.12                       & 0.37             \\ \hline
DER9  & 860 (C)    & 67.5      & 55.5                           & 18                     & 30  & 30            & 0.84               & 0.68                    & 0                       & 0.59             \\ \hline
DER10 & 862 (B)     & 105       & 90                             & 23                     & 90  & 25         & 1.08               & 0.86                    & 0.14                       & 0.51             \\ \hline
\end{tabular}

    \label{tab:DER_Stuff}
\end{table*}
\subsection{Combined Dispatchable and Spinning Reserve Constraints:}

The energy availability factor is a constraint when determining the Dispatchable Factor (DF) and Spinning Reserve Factor (SRF). The sum of the maximum dispatchable power and minimum spinning reserve power equals the total available power. In other words, the sum of DF and SRF should be equal to AF as follows


\begin{equation}\label{equ: inverter_POS}
    \begin{cases}
     DF_1 \ + \ SRF_1 \ = \ AF_1   \\
     DF_2 \ + \ SRF_2 \ = \ AF_2   \\
\vdots\\
    DF_n \ + \ SRF_n \ = \ AF_n
    \end{cases}       
\end{equation}
However, if energy availability exceeds the rated inverter capacity due to the larger size of storage devices, this availability factor may be larger than one. In this case, total dispatchable power and reserve power shouldn't exceed the inverter rated power. Under this scenario, the inverter and storage devices' life are also maximized. In this instance, the inverter power is equal to the summation of dispatchable power and reserve power. 

\begin{equation}\label{equ:inv_RP}
\begin{aligned}
      UCP + SRP = P_{inv}^{ rated} 
\end{aligned}
\end{equation}
which implies, in this condition, AF = 1 and,
\begin{equation}\label{equ:inv_DF_SRF}
\begin{aligned}
      DF_P+SRF_P=1. 
\end{aligned}
\end{equation}

\subsection{Combined Arbitrage Factor and Non-dispatchable Constraints}
The arbitrage and non-dispatchable power operate only in the ESD charging mode. If the DER owners directly charge the BTM' ESD from PV panels during the day time, then the sum of non-dispatchable and arbitrage power should equal PV generation power.
\begin{equation}\label{equ:inv_SAF}
\begin{aligned}
      NDF_p + SAF_p =\frac{P_{pv,p}^{pred}} {P_p^{inv}}
\end{aligned}
\end{equation}
This condition is applied during the daytime PV generation and storage charging cycle. However, this operational constraint is not applicable during the nighttime or no-solar situation, and the ESD is arbitraging power through a bidirectional grid.

\section{Results and Case Studies}

The proposed optimization is implemented on the modified IEEE 34 bus system with ten DERs in ten different buses. Table \ref{tab:DER_Stuff} shows the characteristics of each DER and ESD, as well as the bus that it modifies. Each node is under its constraints as defined in section \ref{sec:con}. We have created a day ahead power flow optimization that spanned 24 hours in 15 minutes increments using that information. It calculates the AF based on the predicted solar PV and storage capacity and current status. It calculates the state of charge, the amount of spinning reserve power available, as well as the unit committed and non-dispatchable generation capabilities at any given point in time.
\begin{figure}[h!]
    \centering
    \includegraphics[scale=0.21]{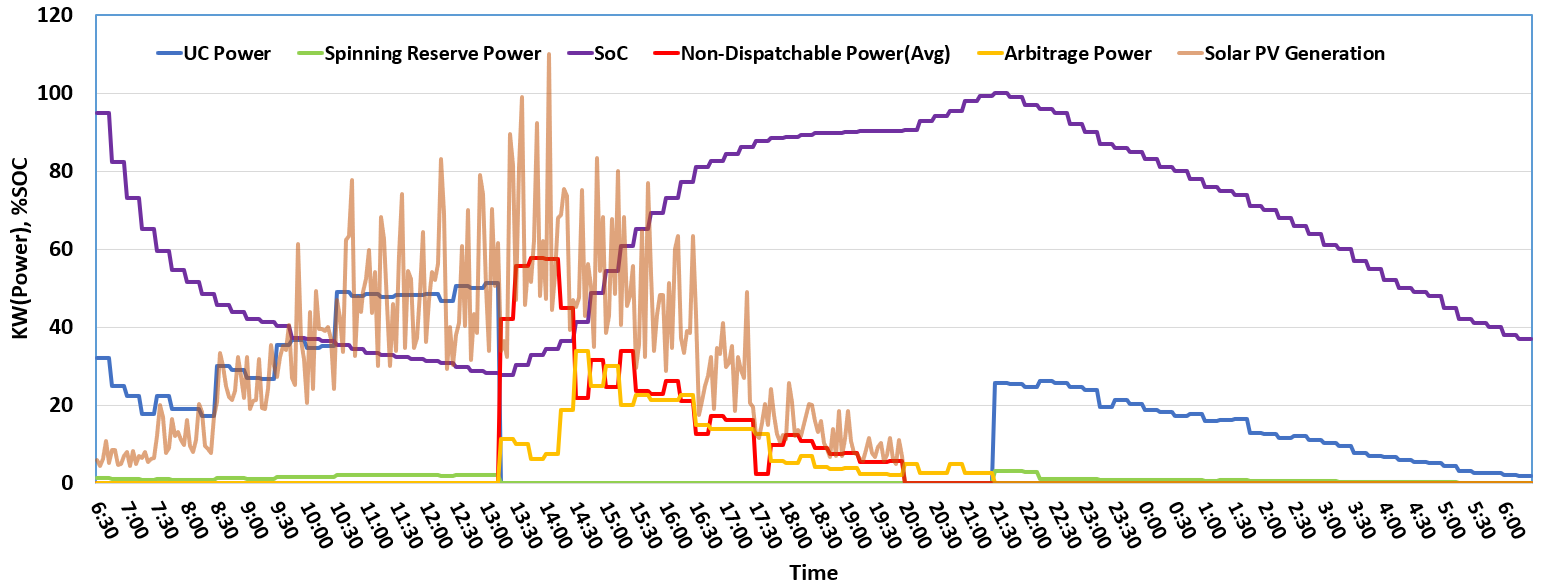}
    \caption{DER1 Different Power Profiles Change Over the 24 Hours Interval.}
    \label{fig:SoC}
\end{figure}

Figure \ref{fig:SoC}  demonstrates how the proposed power profiles change over time for DER1. Based on the different PV generations, storage resources, and energy arbitrage profiles, these power profiles vary over 24 hours of 15 min intervals. The UCP depends on the DER1 storage mode and PV generation profile. During early morning, the UCP is dominated by storage power due to low solar radiation. Gradually, solar PV generation becomes a dominant segment for UCP. Even though solar PV generation fluctuates, the UCP remains the same for a specific interval. The unit commitment is lost when the storage flip to charging mode due to the intermittent nature. In that case, the UCP becomes zero and non-dispatchable or ND power peak due to high solar radiation. ND power decreases to zero due to no PV generation power during the evening. Also, the arbitrage power shows up due to the storage charging mode. This arbitrage power follows the standard storage devices charge profile. In the beginning, the charging rate is low, and arbitrage power is small. Then, it increases and grows steadily before it decreases to the end. When the storage is fully charged around 9:30 PM, it triggers to deliver some unit-committed and spinning reserve power. 
\begin{figure}[h!]
    \centering
    \includegraphics[scale=0.205]{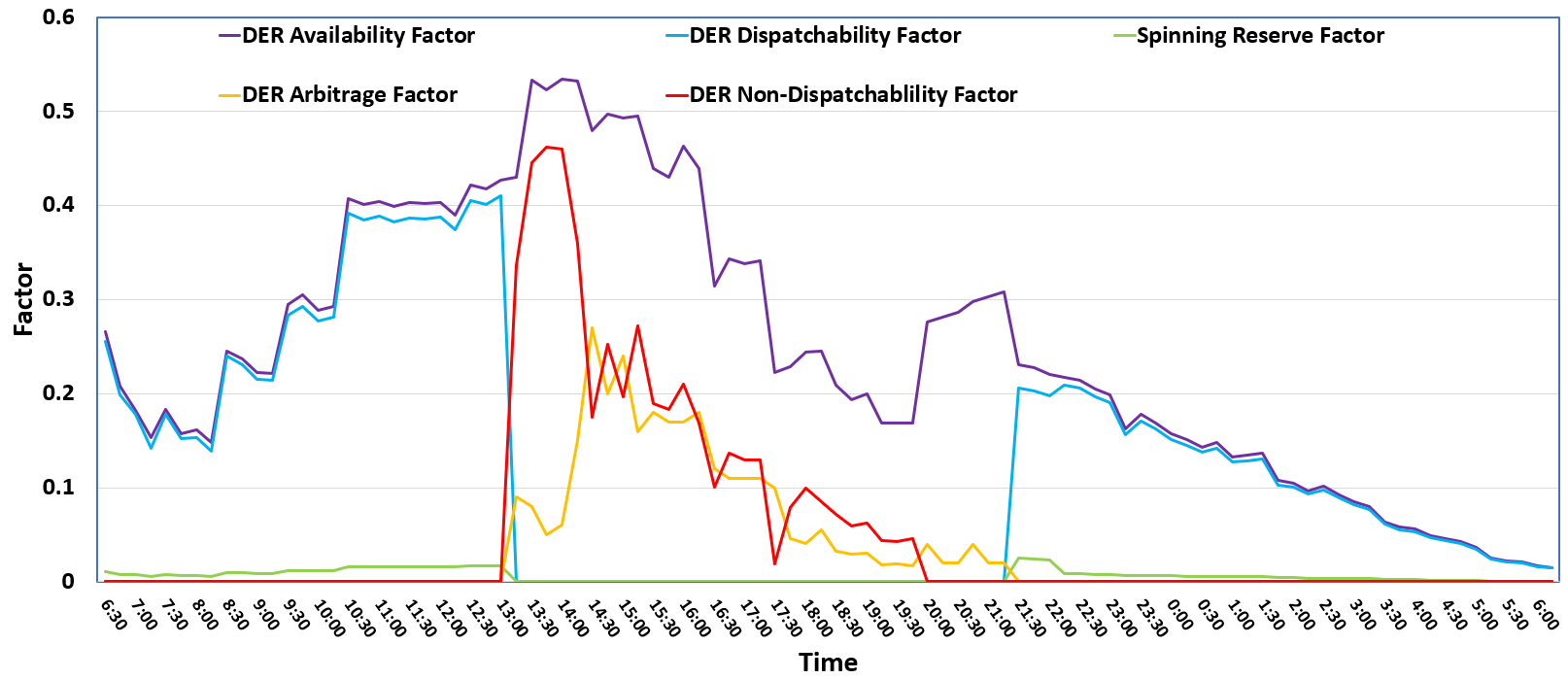}
    \caption{DER1 Operational Factors Change Over the 24 Hours Interval.}
    \label{fig:SoC3}
\end{figure}

Considering the exact PV and storage profile in Fig. 1,  Fig. 2 illustrates the various proposed factors change for DER1 over the 24 hours. AF incorporates the total available power in the PV generation and storage devices. Similar to Fig. 1, DF and SRF continue to impact the DER power profile during the storage discharging mode. However, during the storage charging mode, DF and SRF become zero. Conversely, the NDF and SAF become prominent in this mode. When there is no solar PV generation, the NDF becomes zero. If the ESD is still in charging mode, SAF continues impacting the DER power profile.
Comparing Fig. 1 and Fig. 2, DF follows the exact UCP, SAF follows arbitrage power, NDF follows non-dispatchable power, SRF follows same as spinning reserve power. For rest of this section, we will only show the power profiles as it follows same trend of factors' dynamic profiles. The availability factor represents the sum of PV-generated power and storage power.

The proposed operational matrices are tested on IEEE 34 distribution bus in OpenDSS. Different power profiles (UCP, non-dispatchable power, arbitrage power, spinning reserve power, total DER generated power, and feeder transmitted power) are extracted. This paper considers average solar profiles and loads shapes in summer and winter.

\subsection{Case Study 1: Considering Summer Solar Profile and Load Shape Using Standard Storage Charge/Discharge Rate }
When the storage devices' charging/discharging rate, minimum and maximum SOC are determined by the connected charge controller itself, defined as storage base case. In this condition, the system operator and DER owners can assess all the power profiles under certain initial conditions, as shown in Table 1. Fig. 1 and Fig. 2 only show power profile for DER1, and Fig. 3 and Fig. 4 show power profile for all combined DERs as given in Table 1. 
\begin{figure}[h!]
    \centering
    \includegraphics[scale=0.27]{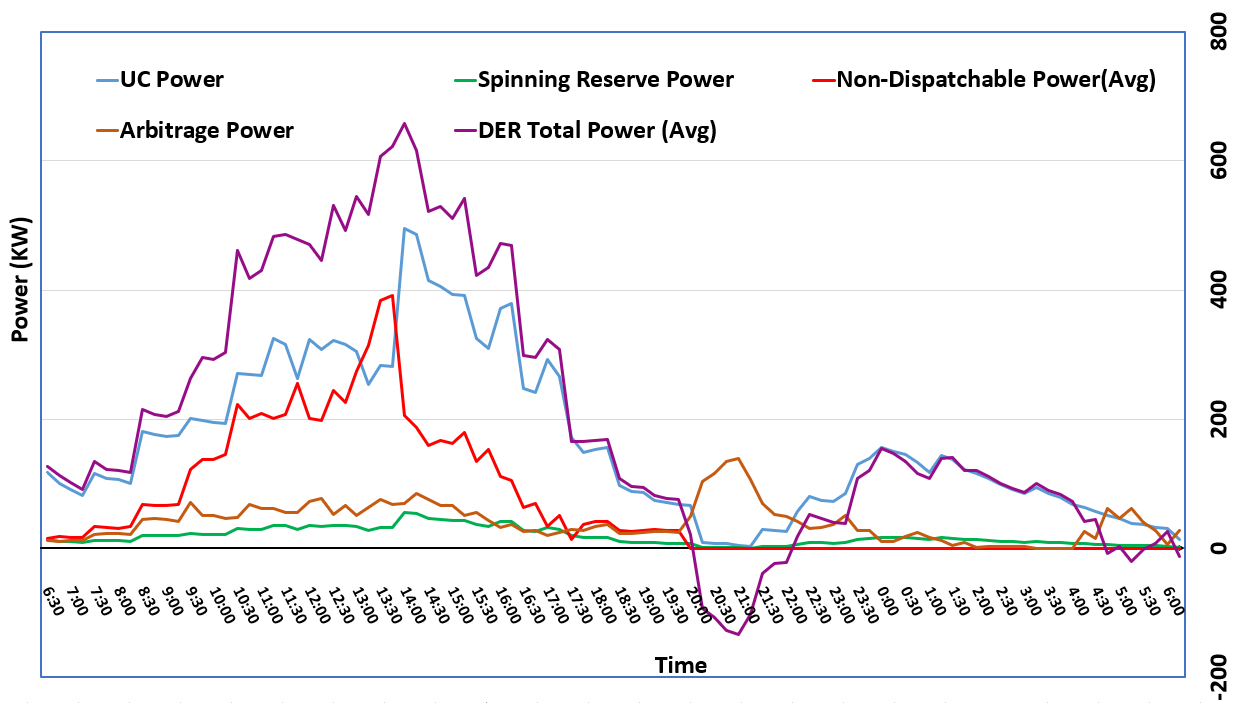}
    \caption{Summer Total DERs Power Profiles}
    \label{fig:sum_der}
\end{figure}

Fig. \ref{fig:sum_der} shows the total amount of unit committed, spinning reserve, and non-dispatchable during the summer time. Indeed, the grid is more stable after providing more UCP, and ESDs' storage cycles are independent of the feeder transfer power.  
According to Fig. \ref{fig:sum_der}, the summer solar profile largely influences the total DER-generated power. According to PJM \cite{pjmlearningcenter}, during the summer, the load is peaked from 17:30 to 20:00. When the DER storage is in the discharging mode, it can provide the UCP. As a result, the UCP dominates the total DER power profile most of the time. However, the non-dispatchable power dominates during noon-time because of high solar radiation. 
We do not consider peak load condition or solar profile in this case. DER's storage devices' charging rate follows the standard multi-stage charging profile \cite{satpathy2020solar}, and storage devices won't discharge until it reaches the maximum SOC. 
\begin{figure}[h!]
    \centering
    \includegraphics[scale=0.27]{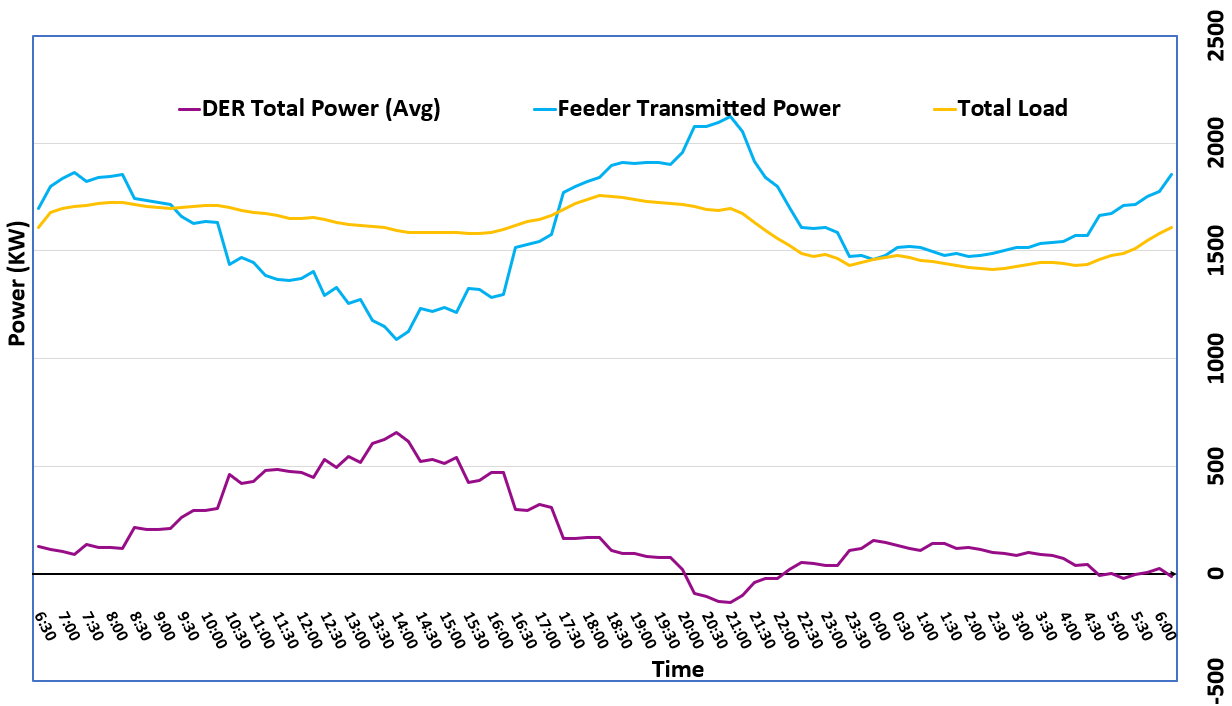}
    \caption{Summer Load, DER, Feeder Transmitted Power Profiles}
    \label{fig:sum}
\end{figure}

Reserve Power is higher when the higher number of ESD is in discharge mode and vice versa. We assume the minimum spinning reserve consumption follows a
Gaussian distribution with a 50\% mean to estimate feeder transmitted power.

Fig. \ref{fig:sum}  illustrates the total DER contribution to the grid for 24 hours interval. This figure also shows the dynamic loads profile, and feeder transfer power during this season. Under this scenario, the DER power profiles change based on the solar PV generation, storage resources, and standard charging/discharging rate of ESD. In such a case, with certain initial condition during the summer load peak time, the DER arbitrage power from the grid could be higher than the DER generated power. In this load peak condition, the equivalent total DER power could be negative and DER would act a load. This power profile configuration suggests that the system operator might need to interrupt the storage charge controller at least during the peak hour.

\subsection{Case Study 2: Considering Winter Solar Profile and Load Shape Using Standard Storage Charge/Discharge Rate}
Fig. 5 illustrate the average power profiles during the winter period. According to PJM, system operator monitor two peaks during this time. Compare to summer, the solar sun hour is much smaller. In this figure, we assume the average winter solar hour from Henderson, Nevada \cite{solarforecast} during this period. We have found two negative DER' equivalent generations without considering the peak load. 
\begin{figure}[h!]
    \centering
    \includegraphics[scale=0.27]{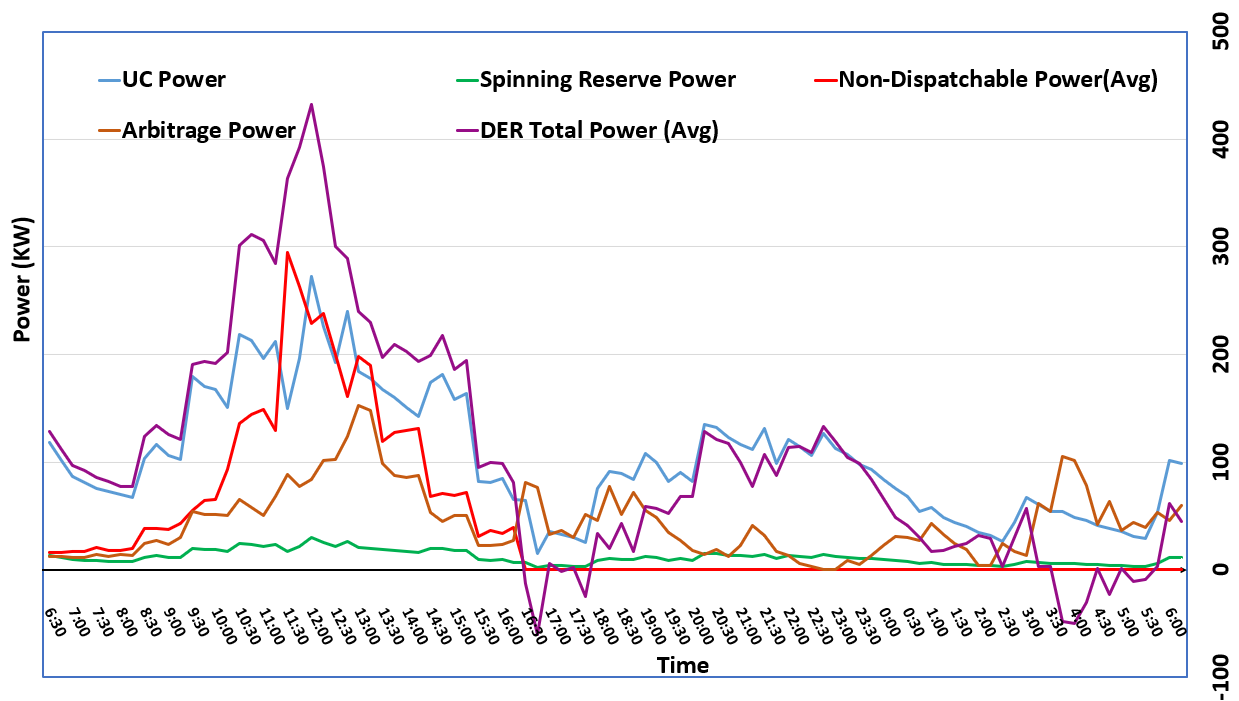}
    \caption{Winter Different DER Power Profiles}
    \label{fig:wint_der}
\end{figure}

In this case study, we consider standard storage charging/discharging rate. Because of that, system operator could not control the DER' different power profiles. However,  system operator would have the information about UCP, Non-dispatchable Power, Arbitrage Power, Spinning Reserve Power and number of units coming from the feeder. Arbitrage power peak during the solar peak radiation. However, according to Fig. 5, we monitor second peak during the evening time.
 
\begin{figure}[h!]
    \centering
    \includegraphics[scale=0.27]{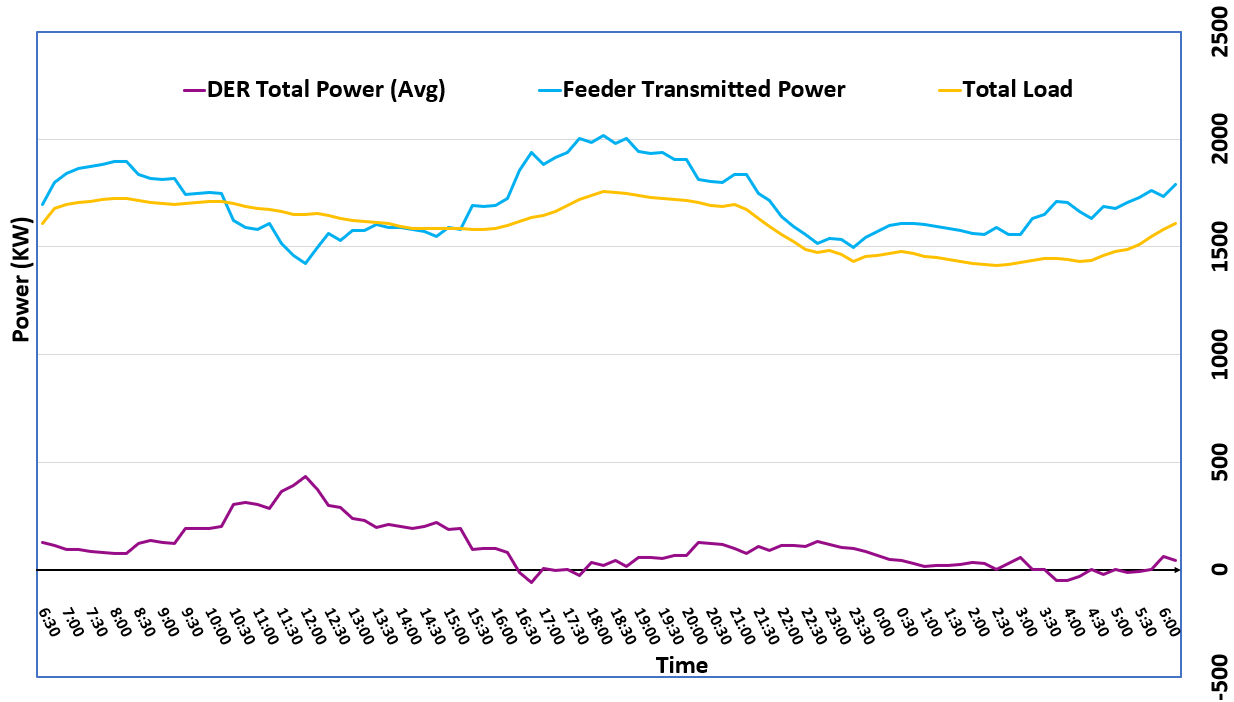}
    \caption{Winter Load, DER, Feeder Transmitted Power Profiles}
    \label{fig:win}
\end{figure}

In Fig. 6, the total DER generation and feeder transmitted power are illustrated for average winter load conditions by using the table 1 initial conditions. From this figure, we able to see two peak load hours in two different time intervals. One happen in the morning 7:30-9:30am, other in the 16:30-19:00 time interval. According to this figure, the equivalent DER generated power become negative during the evening peak load condition because of high arbitrage power. 

\section{Conclusion and Future Works}

This paper has proposed several operational factors to ascertain all  resource and operational constraints from the distributed generator when incorporating ESD to dispatch UCP and SRP. First, considering the ESD's maximum discharge rate, we determine AF to assess the maximum dispatchable available power. Then, to determine the percent UCP and SRP of this AF during the storage discharge mode, we propose DF and SRF. After deciding the percent maximum-minimum range of individual DER' SRF, we maximize the UCP for an optimum total SRP from a particular distribution system using an LP optimization algorithm. Finally, we propose the SAF and NDF to determine the arbitrage and non-dispatchable power during the storage charging mode.  Using these factors, constraints, and LP optimization algorithm, we would be able to extract the individual DER or cluster of DERs with different power profiles like UCP, NDP, SRP, Arbitrage power. These power profiles will help the DER owner enter the electricity market, and the ISO operator picks a particular set of DERs from others in a specific interval. This paper also confirms that the system operator can control the UCP in a particular interval after managing the arbitrage factor, discharging rate, spinning reserve power in the previous intervals. Using this methodology, this research has shown that one can easily and quickly optimize and predict the behavior of a large number of DERs, even within a few minutes for a full day's of operations. This paper also outlines how these factors and power profiles change considering different solar shapes (summer and winter) using the same ESD resources and PV size. Using the standard charging and discharging profile of ESD, this paper illustrates the base case DER power profiles for winter and summer.      

In this paper, we only determine the maximum UCP using the maximum discharge rate of ESD but not incorporating the DER electricity market. Future research will focus on extracting the specific number of UCP after determining the optimal discharge rate using the electricity market. Also, the ESD charging rate would be determined by the arbitrage market. Even though this paper has proposed these operational factors only for individual DER in the distribution system, from the case studies, it is evident that we can apply these operational factors for a cluster of DER or all connected DER. When we combine all of the DER's different powers, we can define the proposed operational factors in a specific distribution system combining all resources.  Future research will propose these operational factors for all distribution and transmission systems, eventually, for the whole power system to understand all the power profiles.

\section*{Acknowledgement}
The work in this paper is supported in part by the U.S. DoE's Office of EERE under the Solar Energy Technologies Office Award Number DE-EE0009022. The views expressed herein do not necessarily represent the views of the DOE or the U.S. Government.

\bibliographystyle{ieeetr}
\bibliography{reference}

\begin{thebibliography}{10}

\bibitem{initiative2012managing}
M.~E. Initiative {\em et~al.}, ``Managing large-scale penetration of
  intermittent renewables.",'' 2012.

\bibitem{7592421}
I.~Abdou and M.~Tkiouat, ``Unit commitment problem in electrical power system:
  A literature review.,'' {\em International Journal of Electrical \& Computer
  Engineering (2088-8708)}, vol.~8, no.~3, 2018.

\bibitem{denholm2010role}
P.~Denholm, E.~Ela, B.~Kirby, and M.~Milligan, ``Role of energy storage with
  renewable electricity generation,'' tech. rep., National Renewable Energy
  Lab.(NREL), Golden, CO (United States), 2010.

\bibitem{yang2018battery}
Y.~Yang, S.~Bremner, C.~Menictas, and M.~Kay, ``Battery energy storage system
  size determination in renewable energy systems: A review,'' {\em Renewable
  and Sustainable Energy Reviews}, vol.~91, pp.~109--125, 2018.

\bibitem{arabali2016new}
A.~Arabali, B.~Asghari, and R.~Sharma, ``A new co-optimization model for grid
  scale storage units in energy and frequency regulation markets,'' in {\em
  2016 IEEE/PES Transmission and Distribution Conference and Exposition
  (T\&D)}, pp.~1--5, IEEE, 2016.

\bibitem{wagner1997large}
R.~Wagner, ``Large lead/acid batteries for frequency regulation, load levelling
  and solar power applications,'' {\em Journal of Power Sources}, vol.~67,
  no.~1-2, pp.~163--172, 1997.

\bibitem{4181668}
T.~S. Dillon, K.~W. Edwin, H.-D. Kochs, and R.~J. Taud, ``Integer programming
  approach to the problem of optimal unit commitment with probabilistic reserve
  determination,'' {\em IEEE Transactions on Power Apparatus and Systems},
  vol.~PAS-97, no.~6, pp.~2154--2166, 1978.

\bibitem{6038963}
B.~Palmintier and M.~Webster, ``Impact of unit commitment constraints on
  generation expansion planning with renewables,'' in {\em 2011 IEEE Power and
  Energy Society General Meeting}, pp.~1--7, 2011.

\bibitem{abujarad2017recent}
S.~Y. Abujarad, M.~W. Mustafa, and J.~J. Jamian, ``Recent approaches of unit
  commitment in the presence of intermittent renewable energy resources: A
  review,'' {\em Renewable and Sustainable Energy Reviews}, vol.~70,
  pp.~215--223, 2017.

\bibitem{remp1949fundamentals}
G.~E. Remp, {\em Fundamentals of power plant engineering}.
\newblock National Press, Millbrae, 1949.

\bibitem{morales2017hidden}
G.~Morales-Espa{\~n}a, L.~Ram{\'\i}rez-Elizondo, and B.~F. Hobbs, ``Hidden
  power system inflexibilities imposed by traditional unit commitment
  formulations,'' {\em Applied Energy}, vol.~191, pp.~223--238, 2017.

\bibitem{howlader2020energy}
H.~O.~R. Howlader, O.~B. Adewuyi, Y.-Y. Hong, P.~Mandal, A.~Mohamed~Hemeida,
  and T.~Senjyu, ``Energy storage system analysis review for optimal unit
  commitment,'' {\em Energies}, vol.~13, no.~1, p.~158, 2020.

\bibitem{zakeri2021centralized}
B.~Zakeri, G.~C. Gissey, P.~E. Dodds, and D.~Subkhankulova, ``Centralized vs.
  distributed energy storage--benefits for residential users,'' {\em Energy},
  vol.~236, p.~121443, 2021.

\bibitem{wu2019safety}
X.~Wu, K.~Song, X.~Zhang, N.~Hu, L.~Li, W.~Li, L.~Zhang, and H.~Zhang, ``Safety
  issues in lithium ion batteries: Materials and cell design,'' {\em Frontiers
  in Energy Research}, vol.~7, p.~65, 2019.

\bibitem{7754038}
J.~W. Shim, G.~Verbič, K.~An, J.~H. Lee, and K.~Hur, ``Decentralized operation
  of multiple energy storage systems: Soc management for frequency
  regulation,'' in {\em 2016 IEEE International Conference on Power System
  Technology (POWERCON)}, pp.~1--5, 2016.

\bibitem{6612760}
F.~Marra, G.~Yang, C.~Træholt, J.~Østergaard, and E.~Larsen, ``A
  decentralized storage strategy for residential feeders with photovoltaics,''
  {\em IEEE Transactions on Smart Grid}, vol.~5, no.~2, pp.~974--981, 2014.

\bibitem{5616799}
T.~D. Hund, S.~Gonzalez, and K.~Barrett, ``Grid-tied pv system energy
  smoothing,'' in {\em 2010 35th IEEE Photovoltaic Specialists Conference},
  pp.~002762--002766, 2010.

\bibitem{eldahab2016enhancing}
Y.~E.~A. Eldahab, N.~H. Saad, and A.~Zekry, ``Enhancing the design of battery
  charging controllers for photovoltaic systems,'' {\em Renewable and
  Sustainable Energy Reviews}, vol.~58, pp.~646--655, 2016.

\bibitem{pjmlearningcenter}
``How energy use varies with the seasons.''
  \url{https://learn.pjm.com/three-priorities/keeping-the-lights-on/how-energy-use-varies}.

\bibitem{satpathy2020solar}
R.~K. Satpathy and V.~Pamuru, {\em Solar PV Power: Design, Manufacturing and
  Applications from Sand to Systems}.
\newblock Academic Press, 2020.

\bibitem{solarforecast}
``Solar forecast {Arbiter API} documentation,'' 2019.
\newblock Observation ID: a8a630c2-99fc-11e9-9ba0-0a580a8200c9.

\end{thebibliography}

\end{document}